\documentclass[10pt,aps,prl,twocolumn,groupedaddress,showkeys]{revtex4-2}
\usepackage{amsmath}
\usepackage{graphicx}
\usepackage[colorlinks=true,linkcolor=blue,urlcolor=blue,citecolor=blue,anchorcolor=blue]{hyperref}
\DeclareGraphicsExtensions{.pdf,.eps,.png,.jpg,.mps}
\usepackage{epstopdf}
\usepackage{threeparttable}
\usepackage{color}
\usepackage[T2A]{fontenc}
\usepackage[utf8]{inputenc}
\usepackage{float}
\usepackage{tabularx}
\usepackage{array}

\begin{document}

\title{Enhanced Piezoelectricity in Sustainable-by-design Chitosan Nanocomposite Elastomers for Prosthetics, Robotics, and Circular Electronics}
\author{Jacopo Nicoletti$^1$}
\author{Leonardo Puppulin$^{1}$}
\email[]{leonardo.puppulin@unive.it}
\author{Julie Routurier$^2$}
\author{Saimir Frroku$^1$}
\author{Nouha Loudhaief$^1$}
\author{Claudia Crestini$^1$}
\author{Alvise Perosa$^1$}
\author{Maurizio Selva$^1$}
\author{Matteo Gigli$^1$}
\email[]{matteo.gigli@unive.it}
\author{Domenico De Fazio$^{1}$}
\author{Giovanni Antonio Salvatore$^1$}
\email[]{giovanni.salvatore@unive.it}
\affiliation{$^1$ Ca’ Foscari University of Venice, Department of Molecular Science and Nanosystems, Via Torino 155, 30172 Venezia, Italy}
\affiliation{$^2$ Université de Haute-Alsace (UHA), Ecole de Chimie de Mulhouse (ENSCMu), 3 Rue Alfred Werner 68200 Mulhouse, France}


\begin{abstract}
Piezoelectricity, the generation of electric charge in response to mechanical stress, is a key property in both natural and synthetic materials. This study significantly boosts the piezoelectric response of chitosan, a biodegradable biopolymer, by integrating chitin/chitosan nanocrystals into natural chitosan-based thin film elastomers. The resulting materials achieve d$_{33}$ values of 15-19 pmV$^{-1}$, a marked improvement over the 5-9 pmV$^{-1}$ observed in pure chitosan films thanks to increased crystallinity from the nanocrystals. We utilize piezoresponse force microscopy (PFM) to accurately measure the d$_{33}$ coefficient, employing an engineered extraction method that eliminates the electrostatic contribution, which can overestimate the piezoelectric response. The resulting chitosan elastomers exhibit elastic deformation up to 40\% strain and a Young's modulus of approximately 100 MPa, similar to soft tissues. These properties, along with the fact that the employed materials can be entirely crafted from upcycled biowaste, make these elastomers ideal for prosthetics, wearable devices, energy harvesters, and sustainable transducers. Our findings underscore the potential of chitosan-based piezoelectric materials for advanced applications in biotechnology, soft robotics, and the green Internet of Things. 
\end{abstract}

\maketitle

\section{\label{Intro}Introduction}
Piezoelectricity is the ability of certain materials to generate an electric charge in response to mechanical stress. It arises from an asymmetric crystal structure that polarizes the material when subjected to mechanical deformation. Piezoelectricity is a phenomenon widely observed in nature, particularly in plants and animals\citep{JosephIEEE2018,SunCRCU2022,SunSCPMA2020,GuerinNM2018,KholkinACSN2010}. In plants, piezoelectric properties have been identified in components such as silk\citep{JosephIEEE2018} and cellulose nanocrystals\citep{SunCRCU2022}, contributing to their structural integrity and responsiveness to mechanical stimuli. Similarly, animals employ piezoelectricity in biological processes, with collagen in connective tissues and bones exhibiting notable piezoelectric behaviour crucial for functions such as bone remodelling and repair\citep{SunSCPMA2020}. Proteins and their derivatives (such as glycine crystals) and peptides (such as diphenylalanine) exhibit notable piezoelectric characteristics\citep{GuerinNM2018}, especially in self-assembled nanotubes\citep{KholkinACSN2010}. The M13 phage virus, with its helical protein-clad filamentous structure, can be used to create highly ordered crystalline structures\citep{ShinEES2015}. Ultimately, the epidermis of living human skin possesses a continual electric dipole moment oriented perpendicular to its surface, which reacts to mechanical stimuli, generating electrical signals transmitted to the neural system\citep{AthenstaedtS1982}.

The convergence of piezoelectric materials with micro and nanotechnologies has sparked revolutions in biotechnology, bioelectronics, and healthcare. From wearable sensors\citep{AliAHM2023} for personalised health monitoring\citep{SultanaJMCB2017} to intricate actuators orchestrating precise cellular manipulations\citep{DuJCIS2020}, these innovations showcase the transformative potential of piezoelectricity in enhancing human well-being. As we delve deeper into the realms of biomedicine and healthcare, it is crucial for piezoelectric materials to prioritise biocompatibility and mechanical flexibility, often necessary for specific applications\citep{KamelBR2022,WuMD2021}. Furthermore, the pervasive use of these devices demands materials that are eco-friendly but also amenable to large-area processing and thin film fabrication\citep{LemaireSMS2018}.

A common metric to quantify piezoelectricity is the d$_{33}$ coefficient which measures the amount of electric charge generated per unit of mechanical stress applied perpendicular to the electric field. A higher d$_{33}$ coefficient indicates a more robust piezoelectric response, enhancing energy conversion efficiency. While traditional materials such as lead zirconate titanate (PZT) and barium titanate (BTO) possess large piezoelectric response (200-600 pmV$^{-1}$)\citep{KarakiJJAP2007,SmithJACS2012,GaoA2017}, they are mechanically rigid\citep{FanAM1999}, brittle\citep{VandenEndeJMS2007}, and toxic\citep{PandaJMS2009,AhamedN2020}. Moreover, their production and disposal processes are far from being environmentally friendly\citep{PandaJMS2009}. Among synthetic piezoelectric polymers, polyvinylidene fluoride (PVDF) is widely used for flexible transducers\citep{GuanACSAMI2023,StiubianuM2022,TuguiJMCC2017,DeMarzoAEE2023}; however, it generates fluorine waste, leading to significant environmental concerns\citep{VeigaMTC2020}. Table \ref{tab:Tab1} in Supplementary Information (SI) contains a list of soft synthetic piezoelectric polymers. Therefore, increasing interest has been directed towards the study of natural bio-based piezoelectric materials\citep{HanninenCP2018}. These materials show well-organized structures and low symmetrical patterns, enabling efficient electromechanical interactions\citep{AliAHM2023,KumarG2024}. Table \ref{tab:Tab2} in Supplementary Information (SI) contains a list of piezoelectric biopolymers. 

Our interest has focused on chitosan, a cost-effective natural biopolymer derived from chitin, which is the second most abundant polysaccharide on Earth\citep{MaschmeyerCSR2020,XuCSR2019}. This material offers numerous advantages such as biocompatibility\citep{RodriguesJFB2012}, biodegradability\citep{WronskaF2023}, and the ability to easily form thin films\citep{XuCSR2019,PavinattoB2010}. Chitosan can be extracted as a food industry byproduct from the exoskeletons of crustaceans\citep{HamedTFST2016}, therefore it is considered “sustainable by design”. In this regard, numerous studies focus on developing methods that prioritize eco-friendly practices for its acquisition and processing\citep{ShamshinaACSSC2016}. These desirable properties make chitosan highly suitable for a range of industrial and biomedical applications, including tissue engineering\citep{SultankulovB2019}, wound healing\citep{ChenAMT2020}, and optical sensors\citep{FenIEEESJ2013,LinBC2012}, due to the transparency of the resulting films\citep{ParkPNAS2021}. The study of the electromechanical properties of this biopolymer started decades ago\citep{FukadaJPS1975} with more recent investigations focusing on thin films\citep{AhmadIOPCS2020,PraveenRSCA2017}. However, due to the amorphous portion of the film matrix, the piezoelectric properties of pure chitosan appear weak\citep{HanninenPE2016}. The d$_{33}$ coefficient of pristine thin film chitosan is, indeed, in the range of 5-9 pmV$^{-1}$\citep{HanninenCP2018,ToalaIJBM2023}. One strategy to enhance the piezoelectric response consists in the formulation of chitosan-based composite materials\citep{ToalaIJBM2023}. Recent studies show that increasing the poly-hydroxybutyrate (PHB) content in chitosan films enhances crystallinity, and, as a consequence, the d$_{33}$\citep{ToalaIJBM2023}. Another study successfully crafted a flexible piezoelectric pressure sensor using biodegradable glycine and chitosan film, leveraging chitosan's flexibility and glycine's piezoelectric potential\citep{HosseiniACSAMI2020}. Additionally, blending chitosan with nanocellulose structures like cellulose nanofibrils\citep{QinPT2023} or cellulose nanocrystals\citep{YadavP2020} hints potential paths for enhancing the piezoelectric performance. Other examples include the use of chitin or chitosan nanofibers with poly(vinylidene fluoride) (PVDF)\citep{HoqueJMCA2018} and H$_2$SO$_{4}$-treated PEDOT\citep{DuJCIS2020}. One breakthrough in the field occurred in 2023 when Ref.\citenum{DeMarzoAEE2023} reported d$_{33}$ values up to 15 pmV$^{-1}$ by optimising the neutralisation conditions of pristine chitosan thin films in an alkaline environment. Table \ref{tab:Tab3} in SI lists various chitosan-based materials categorised by the production methodology, the d$_{33}$ measurement technique and values obtained.   

Despite the extensive research on chitosan thin films, the piezoelectricity of all-natural chitosan elastomers has not been investigated so far. Piezoelectric materials that can elastically deform are of great interest in wearable skin-like devices for health monitoring\citep{HuNBM2023,ShuR2021,SunNBE2020}, energy scavenging\citep{DagdevirenEML2016,DuNR2020}, and actuation\citep{DagdevirenEML2016,DuNR2020}. Here, we report a study of the nanoscale piezoelectricity of all-natural elastomers prepared from chitosan, glycerol and chitin/chitosan nanocrystals. We achieved d$_{33}$ in the range of 15-19 pmV$^{-1}$ for the chitosan elastomers, a more than two-fold improvement compared to the d$_{33}$ of 5-9 pmV$^{-1}$ found in pure chitosan films. The enhanced piezoelectric response is explained by an increased crystallinity thanks to the addition of natural biobased chitosan/chitin NCs. Our accurate extraction of the d$_{33}$ in soft elastomers entails two critical challenges: (i) achieve high sensitivity of the piezoresponse force microscopy (PFM) transduction without damaging the surface and (ii) extract the effective d$_{33}$ that should not be affected by the electrostatic coupling between the cantilever and the surface. We have ensured a proper matching of the spring constant of the cantilever with the polymer surface (essential to achieve the desired sensitivity and avoid surface indentation). Moreover, we have refined a methodology for the extraction of the d$_{33}$ coefficient to depurate the measurements from the electrostatic interaction that may completely override the piezoelectric response, hampering the measurement\citep{MillerNA2019,Gervacio-ArciniegaJAP2020}. This novel approach avoids tedious calibration procedures against reference samples. The content of NCs has been optimized to achieve elastic deformation up to 40\% of strain and Young's modulus of about 100 MPa, aligning closely with the physiological modulus of soft tissues. This opens up promising avenues for the construction of skin-mounted transducers, sensors, and energy scavengers that are both biocompatible and sustainable by design, as well as for the realisation of green Internet of Things devices. 

\section{\label{Results}Results and Discussion}

\subsection{\label{Film}Film Preparation and Characterisation}

\begin{figure*}[htbp!]
\centerline{\includegraphics[width=180mm]{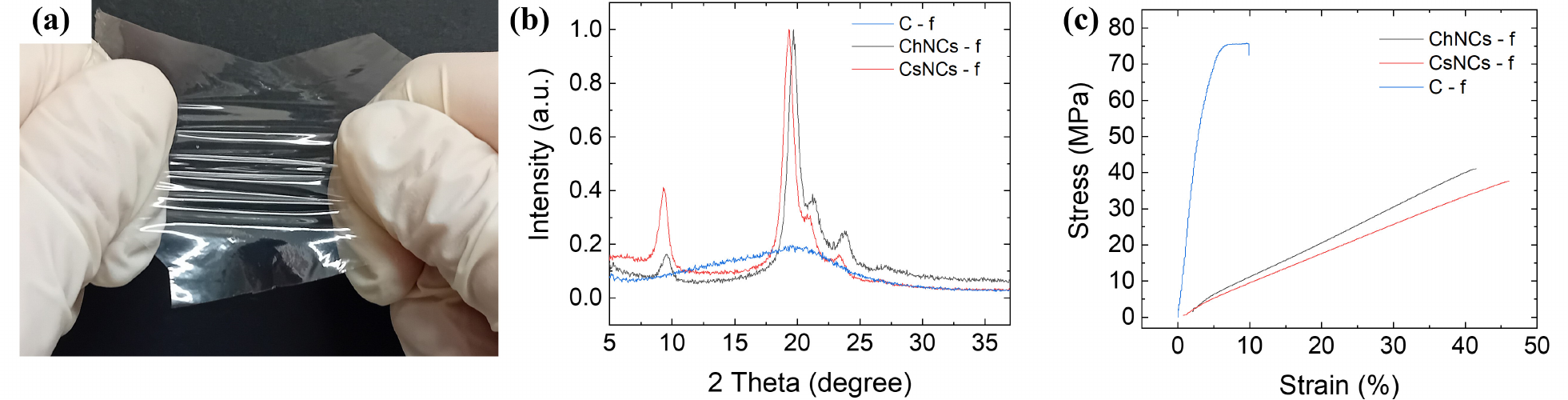}}
\caption{(a) Optical image of a thin film composed of chitosan, glycerol, and chitin nanocrystals, exhibiting transparency, flexibility, and stretchability. (b)  XRD pattern of films composed of pure chitosan (C - f blue curve); chitosan, glycerol, and chitin nanocrystals (ChNCs - f, black curve) with 40-40-20 wt\% composition; chitosan, glycerol, and chitosan nanocrystals (CsNCs - f, red curve) with 40-40-20 wt\% composition. c) Stress/strain curves of the former films with the same colour coding.}
\label{fig:Fig1}
\end{figure*}

Chitin nanocrystals (ChNCs) and chitosan nanocrystals (CsNCs) were synthesized as reported in the literature with minor modifications\citep{MassariUR,SegatoUR}. The details of the production process are described in the Experimental section.  The ChNCs and the CsNCs were characterized from the dimensional and structural point of view. Dynamic light scattering (DLS) analysis highlights a length of 350 and 300 nm, for ChNCs and CsNCs, respectively, and a Z-potential of about +55 mV for both, in agreement with literature data\citep{PereiraCP2015}. FT-IR analyses, previously reported\citep{MassariUR,SegatoUR}, do not show significant differences between the two samples, confirming that the deacetylation step does not introduce additional structural modifications. Lastly, the degree of deacetylation ($DD$) is calculated as follows\citep{DosSantosCR2009}:

\begin{equation}
\label{eq:Eq1}
DD = 100 \times (4-0.583093\times W_{C/N})
\end{equation}

where $W_{C/N}$ is the weight carbon to nitrogen ratio, as determined by CHNS analysis. ChNCs and CsNCs are found to have values of 4.6\% and 36.8\%, respectively. Three different film compositions were selected for the study: pure chitosan (C – f) and two nanocomposite formulations containing chitosan (40 wt\%), glycerol (40 wt\%), and either chitin or chitosan nanocrystals (20 wt\%) (ChNCs – f, and CsNCs – f respectively). Films  were cast on various substrates including glass and Kapton foils using an automatic applicator that allows to achieve good uniformity and reproducibility. For a full overview of the production methods of our films see the Experimental Section. As shown in Figure \ref{fig:Fig1}(a), our fabrication method yields transparent (see Figure \ref{fig:FigS1} in SI for the optical absorption measurement), flexible, and stretchable thin films with thickness in the range of 15-20 µm. These properties are particularly attractive for applications in electronics and optoelectronics\citep{TrungJMCC2017}.

Chitosan-based films normally possess a partially crystalline structure\citep{IoelovichJC2014}. The X-ray diffraction (XRD) pattern of our pure chitosan films (blue line in Figure \ref{fig:Fig1}(b)) appears largely amorphous, with a broad band around 15-25°, indicating modest crystallinity. Conversely, chitosan films containing glycerol and chitin nanocrystals (ChNCs) or chitosan nanocrystals (CsNCs) (black and red lines in Figure \ref{fig:Fig1}(b), respectively) show sharp and narrow diffraction peaks, indicating higher crystallinity. The most notable peaks are observed at 2 theta angles of $\sim$ 9.5° and $\sim$ 21.0°, as reported in previous studies\citep{RoblesCP2016,GoodrichB2007}. We do not observe notable differences between the crystallinity of films containing ChNCs and CsNCs. While it is difficult to accurately quantify the crystallinity of the films, the XRD spectra suggest that the presence of NCs in chitosan films enhances their crystallinity. 

Tensile tests provide information about key mechanical parameters such as maximum load, Young’s modulus, and elongation at break (see Figure \ref{fig:Fig1}(c)). The results of the tests demonstrate that the presence of glycerol alters the stress-strain curve, causing films to exhibit typical elastomer response: a non-linear behaviour with large deformation at low stress and without yield strength point\citep{Callister2008}. The film with pure chitosan can endure a maximum load of $\sim$ 70 MPa and $\sim$ 8\% elongation before breaking. This leads to Young’s modulus of $\sim$ 875 MPa. Combining chitosan, glycerol, and NCs results in films with an excellent compromise between stiffness and deformability with values of elongation at break, maximum load and Young’s modulus being $\sim$ 40\%, $\sim$ 40 MPa, and $\sim$ 100 MPa, respectively. These values interestingly align with those of tissues like tendons and muscles\citep{McKeeTEPBR2011} prospecting applications in prosthetics and soft robotics\citep{AshuriBML2020}. Adjusting the percentage of NCs allows fine-tuning of the mechanical properties for targeted applications. Besides their desirable mechanical properties, optical transparency expands the range of their potential applications to include optical sensors and transducers. 

\subsection{\label{PFM}Piezoresponse Force Microscopy (PFM) Analysis}

\begin{figure*}[htbp!]
\centerline{\includegraphics[width=180mm]{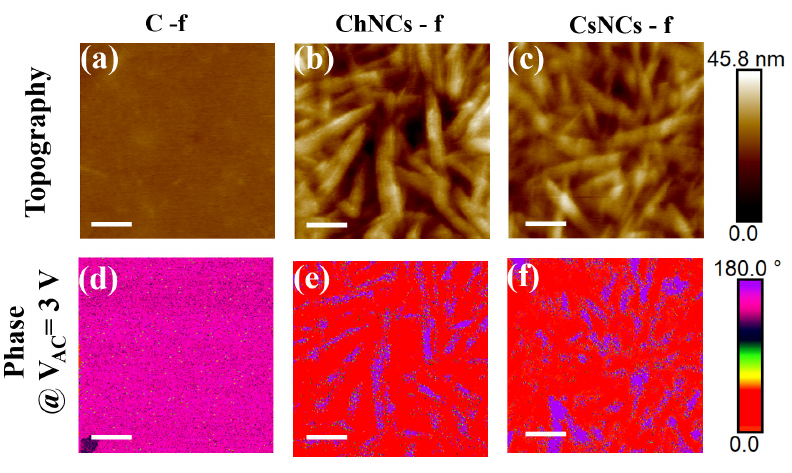}}
\caption{PFM images of the topography (a-c) and phase at $V_{AC}$ = 3 V (d-f) of C - f, ChNCs - f, CsNCs - f. The scale bar is 200 nm in all subplots, while the colour scales on the right refer to each subplot in their respective line.}
\label{fig:Fig2}
\end{figure*}

We used an atomic force microscope in contact mode to examine the properties of our films. Raster scans in the PFM mode were performed over selected areas of the sample surface thus enabling simultaneous acquisition of the topography and out-of-plane PFM phase images. Using a conductive tip with a low spring constant of 0.1 N/m minimizes the risk of damaging the test material and ensures the required sensitivity for precise electromechanical characterization of these soft materials. Analysis of the topography reveals a distinctive texture in films containing nanocrystals, unlike those composed solely of chitosan, as is evident from Figure \ref{fig:Fig2}(a-c). The nanocrystals are uniformly distributed throughout the chitosan matrix and form intricate, intertwined patterns. This arrangement is clearly visible in Figure \ref{fig:FigS2}, where AFM images in tapping mode reveal the topography over a larger scan size of 10 × 10 µm². Roughness measurements also reflect the presence of NCs (see Table \ref{tab:Tab4} in Supplementary Information), yielding values in the range of 3.0 - 5.0 nm compared to the 0.5 nm of pure chitosan (calculated over a 1 µm × 1 µm area on multiple locations of the films).

Out-of-plane PFM phase imaging enables the visualisation and mapping of variations in the piezoelectric response across a sample surface with high spatial resolution. This technique can reveal important insights into the electromechanical behaviour of materials, such as ferroelectric/piezoelectric domains and domain switching behaviour. We use a conductive gold substrate to generate an electric field in the sample, while the conductive PFM cantilever acts as the second electrode. Figure \ref{fig:Fig2}(d-f) provides the results of out-of-plane PFM phase imaging for our films at $V_{AC}$ bias of 3 V. For all measurements, the $V_{DC}$ was adjusted to approximately 0.4 V, a value closely aligned with the common surface electrostatic potential, $V_{EL}$, across all three samples, as detailed in subsequent sections. This DC voltage was selected to significantly reduce electrostatic forces in the cantilever oscillation, thereby emphasizing the piezoelectric response. Under this experimental conditions, oppositely oriented vertical polar domains display an out-of-plane phase contrast of 180°. When probed using $V_{AC}$ = 3 V, pure chitosan films display uniform out-of-plane phase of oscillation across the scanned area (Figure \ref{fig:Fig2}(d)). In contrast, films containing NCs reveal alternating regions with a 180° phase difference, indicative of piezoelectric domains. This phenomenon is particularly pronounced in areas where nanocrystals are present within the film. Given that measurements were taken at a fixed frequency, specifically the cantilever resonance frequency measured at a single point on the sample, the observed phase variation can also be linked to changes in contact forces experienced by the cantilever due to variations in topography and contact stiffness (\textit{i.e.}, matrix \textit{vs} nanocrystals). Nonetheless, the bimodal distribution of phase characterized by two populations separated by 180° suggests that findings in Figure \ref{fig:Fig2}(e-f) represent the initial evidence showcasing the piezoelectric nature of all examined films.

\begin{figure*}[htbp!]
\centerline{\includegraphics[width=180mm]{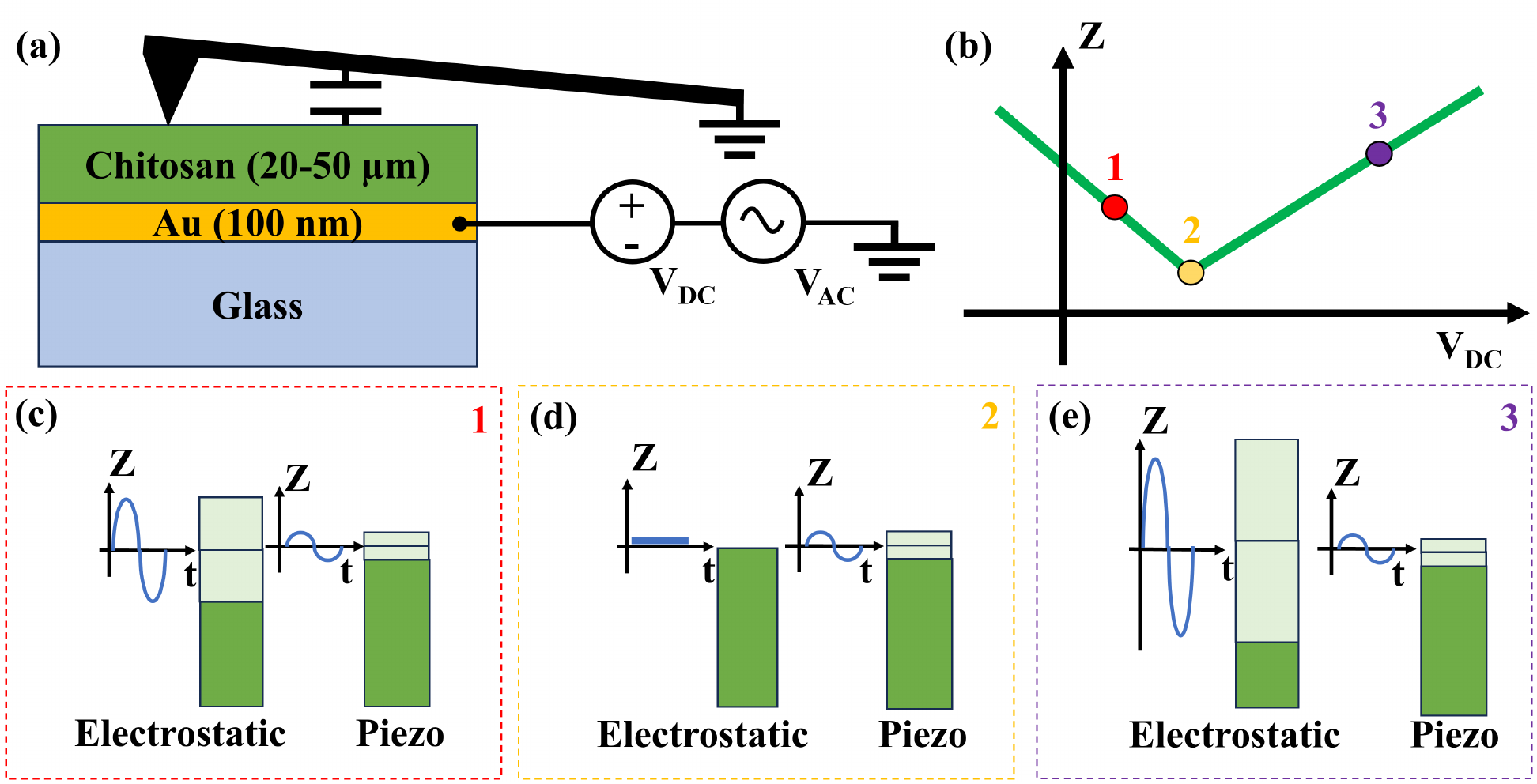}}
\caption{(a) Schematics showing the grounded PFM cantilever capacitively coupled to our sample with an AC and DC voltage applied to the gold electrode. (b) Displacement response of the biopolymer as a function of the applied $V_{DC}$ and (c-e) the unravelling of the two contributions (electrostatic and piezo) to the displacement $Z$ at three representative points. }
\label{fig:Fig3}
\end{figure*}

While PFM imaging is crucial for surface domain visualisation, it lacks the precision necessary for accurate d$_{33}$ coefficient quantification. Typically, surface scans incorporate an electrostatic capacitive component alongside the piezoelectric, posing challenges in its elimination during imaging. As shown in Figure \ref{fig:Fig3}(a), the capacitor represents the capacitive coupling between the cantilever and the sample. When a sinusoidal voltage $V_{AC}$ in addition to a constant voltage $V_{DC}$ is applied to the substrate, a displacement $Z$ is recorded through the cantilever deflection. The displacement $Z$ depends on $V_{DC}$ and $V_{AC}$ through Equation \ref{eq:Eq2}\citep{MillerNA2019}:

\begin{equation}
\label{eq:Eq2}
Z=\text{d}_{33}V_{AC}+\frac{1}{k}\frac{\partial C}{\partial Z}V_{AC}\mid V_{DC}-V_{EL}\mid
\end{equation}

\begin{figure*}[htbp!]
\centerline{\includegraphics[width=120mm]{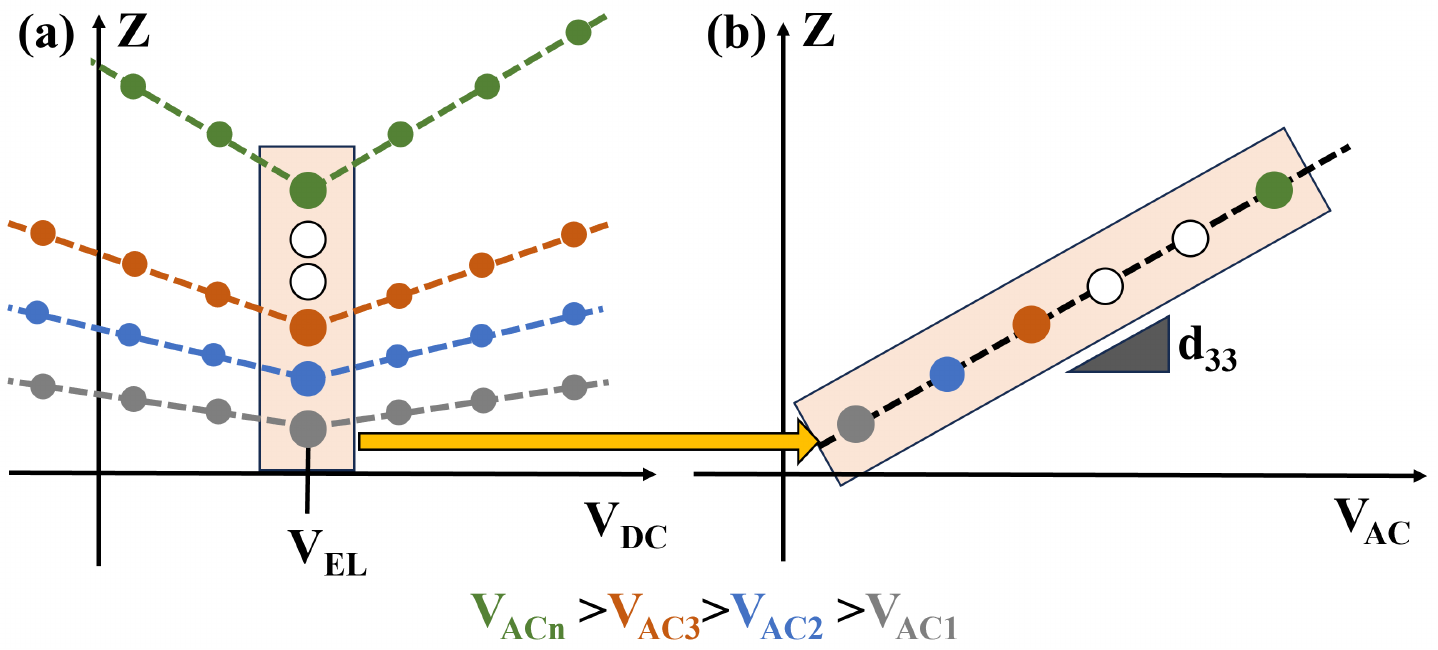}}
\caption{(a) Qualitative $Z$ as a function of $V_{DC}$ at several $V_{AC}$s and (b) a focus on the increase of $Z$ as a function $V_{AC}$ when $V_{DC}$ is set to $V_{EL}$. The slope of the graph (b) can be used to extrapolate the piezoelectric coefficient d$_{33}$ of the material through Equation \ref{eq:Eq2}.}
\label{fig:Fig4}
\end{figure*}

\begin{figure*}[htbp!]
\centerline{\includegraphics[width=180mm]{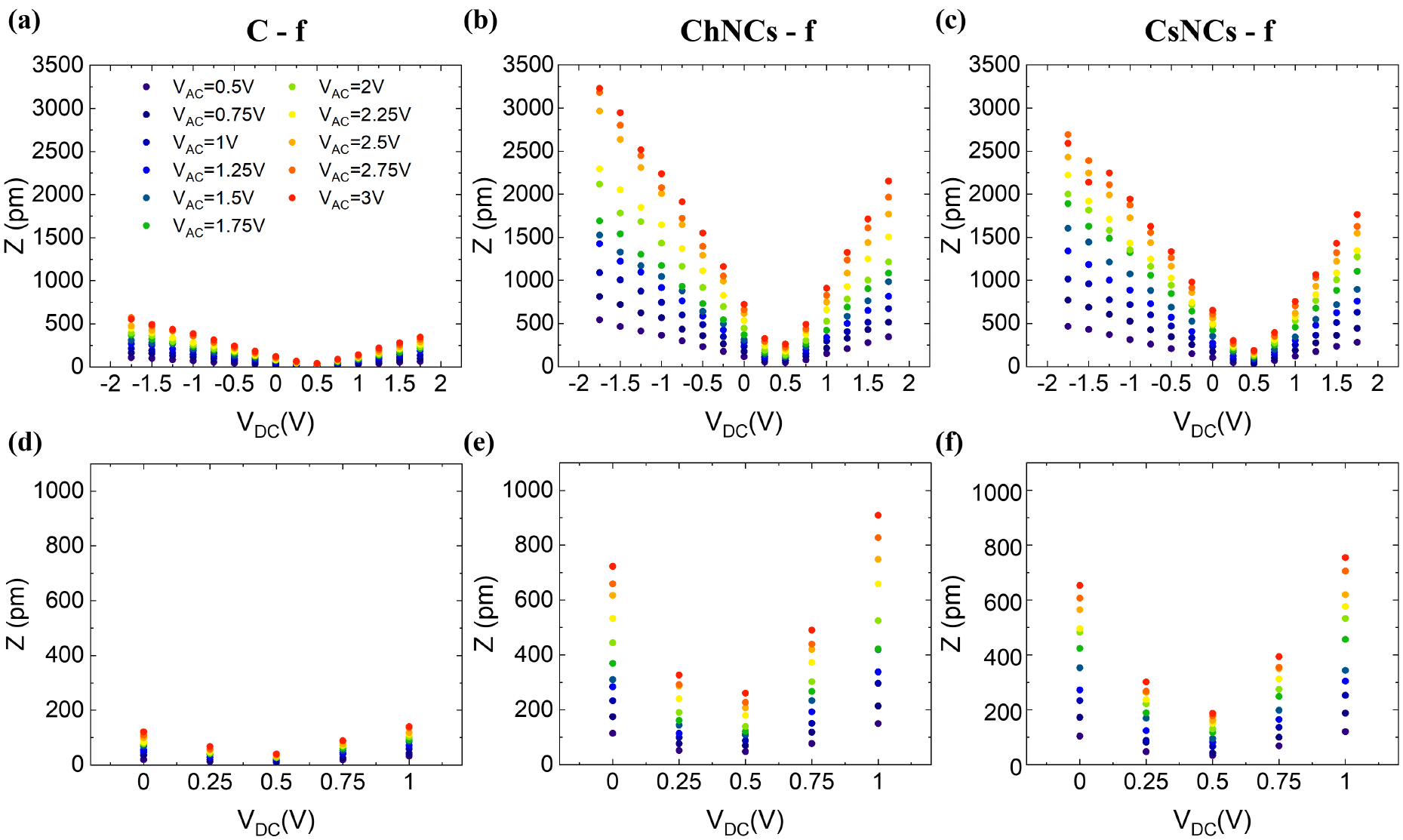}}
\caption{(a-c) Graphs and (d-f) close-ups showing the displacement Z trend as a function of $V_{DC}$ (-1.75-1.75 V, while 0-1 V in the close-ups) at different $V_{AC}$ (0.5-3 V) for chitosan -f (a,d),  ChNCs – f (b,e), and CsNCs - f (c,f). The plots show the data before the normalisation by the quality factor ($Q$) of the cantilever resonance curve.  }
\label{fig:Fig5}
\end{figure*}

here $k$ is the tip spring constant of the cantilever (0.1 Nm$^{-1}$) and $\frac{\partial C}{\partial Z}$ is a term accounting for the tip-sample capacitive coupling, which varies with $Z$. The $Z$ response is the sum of two addends: the first component is related to the piezoelectric contribution, and it is linearly dependent on $V_{AC}$ and independent of $V_{DC}$. The second component is an electrostatic artefact, which grows as the applied $V_{DC}$ voltage moves away from a minimum value, here called $V_{EL}$\citep{MillerNA2019}. Figure \ref{fig:Fig3}(b) depicts the qualitative influence of $V_{DC}$ on $Z$. Point 2 is the minimum of the $Z$ \textit{vs} $V_{DC}$ curve, which is reached at $V_{DC}$ = $V_{EL}$. At this point, the electrostatic coupling between the cantilever and the surface is perfectly counterbalanced by the applied $V_{DC}$, therefore the displacement resulting from electrostatic forces is zero and only the piezoelectric contribution is present. Point 1 and Point 3 are characterised by a larger electrostatic contribution compared to point 2. Figure \ref{fig:Fig3}(c-e)  This is due to a larger distance between $V_{DC}$ and $V_{EL}$ at points 1 and 3 compared to point 2. In this scenario, higher $V_{AC}$ leads to higher $Z$ responses both at $V_{EL}$ and away from it. Therefore, an accurate estimation of the d$_{33}$ coefficient requires the extrapolation of the slope of the Z versus $V_{AC}$ curve (Figure \ref{fig:Fig4}) at $V_{DC}$ = $V_{EL}$. Notably, other reported methods extract the d$_{33}$ at constant $\mid V_{DC}$-$V_{EL}\mid$ values. While this approach reduces errors, it does not eliminate them entirely, as $\frac{\partial C}{\partial Z}$ may still depend on the biasing voltage.

Using PFM, we evaluated the displacement $Z$ as a function of $V_{DC}$ and $V_{AC}$. Specifically, we swept $V_{DC}$ from -1.75 V to 1.75 V with 0.25 V increments, while measuring $Z$. We then repeated the sweep at several $V_{AC}$s ranging from 0.5 V to 3 V with 0.25 V increments. The results of our tests on C - f, ChNCs – f ,  CsNCs - f  are reported in Figure \ref{fig:Fig5}. Seen as a function of $V_{DC}$, each displacement ($Z$) curve has a V-shape. Generally, higher values of $V_{AC}$ produce larger $Z$. The presence of both ChNCs and CsNCs in the films also leads to larger $Z$ for all values of $V_{DC}$, including $V_{DC}$ = $V_{EL}$: this indicates that the presence of nanocrystals induces a greater piezoelectric response.

\begin{figure}[htbp!]
\centerline{\includegraphics[width=80mm]{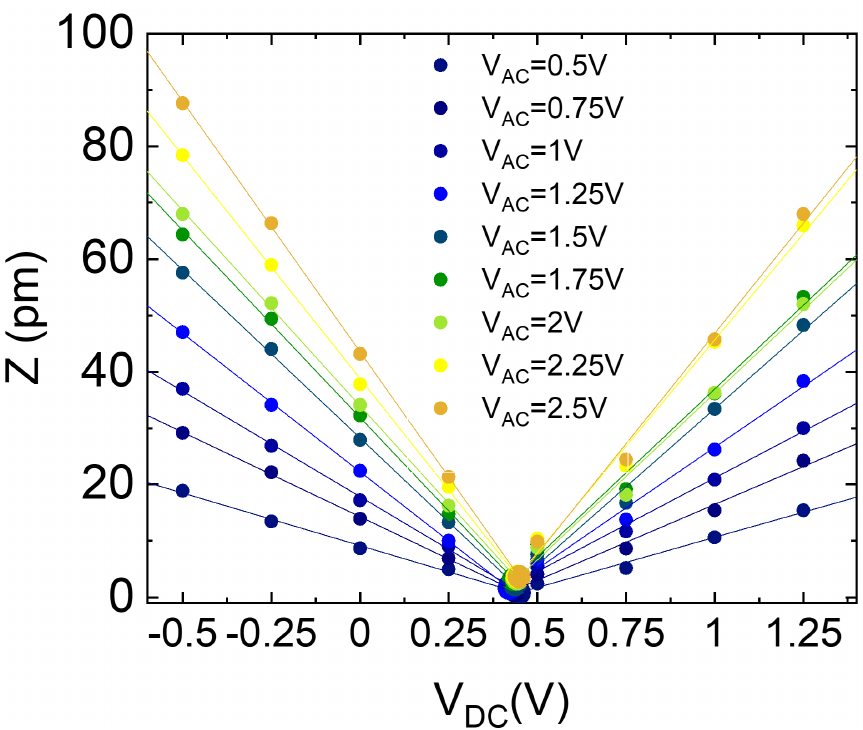}}
\caption{Representative fits of $V_{DC}$ sweeps with two straight lines intersecting at $V_{DC}\sim V_{EL}$ in a  CsNCs - f. In this example $V_{EL} \sim$ 0.4 V for all values of $V_{AC}$.}
\label{fig:Fig6}
\end{figure}

\begin{figure*}[htbp!]
\centerline{\includegraphics[width=180mm]{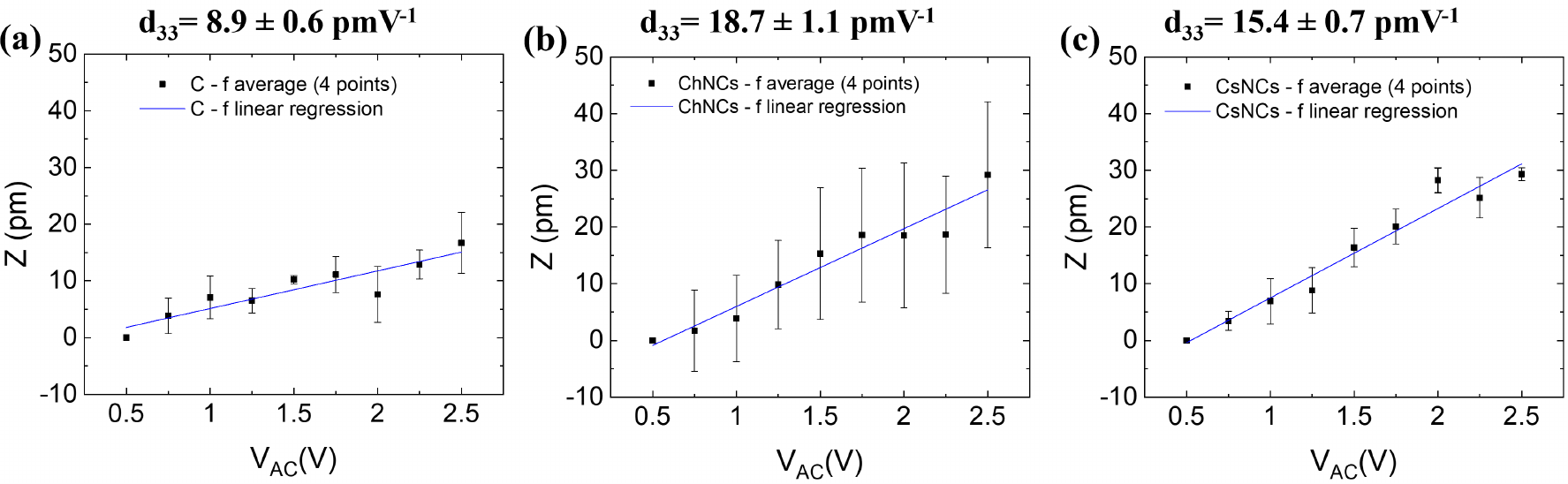}}
\caption{Average piezoresponse $Z$ as a function of V$_{AC}$ fitted linearly for films composed of C - f (a), ChNCs - f (b), and CsNCs - f (c). The slope of the curves, hence d$_{33}$, is printed at the top of the graphs.}
\label{fig:Fig7}
\end{figure*}

To correctly estimate the piezoelectric coefficient d$_{33}$, the $Z$ displacement values shown in Figure \ref{fig:Fig5} need to be normalised by the quality factor $Q$ of the cantilever resonance curve. This normalization also takes into account the amplification factor of the lock-in. To identify the value of $V_{EL}$ at each $V_{AC}$, which could fall in between two measurement points, we fit for each film the displacement $Z$ as a function of $V_{DC}$ and note down the intersection voltage of the descending and ascending slopes as the precise value of $V_{DC}$ (Figure \ref{fig:Fig6}). In this and the following analyses, we excluded the data points corresponding to $V_{AC}$ = 2.75 V and $V_{AC}$ = 3 V as we revealed the presence of non-linear distortion effects.

We then set $V_{DC}$ = $V_{EL}$ and interpolate the experimental points of $Z$ as a function of $V_{AC}$ to extrapolate the d$_{33}$. Measurements were repeated on 5 different points for each sample and at different locations. The graphs in Figure \ref{fig:Fig7} show the mean and standard deviation of all measurements for chitosan (a), chitosan-glycerol-ChNCs (b) and chitosan-glycerol-CsNCs (c). As expected, $Z$ plotted as a function of $V_{AC}$ at $V_{DC}$ = $V_{EL}$ displays linear trends. The slope of the fitting line represents the piezoelectric coefficient d$_{33}$. Each point in Figure \ref{fig:Fig7} averages five different measurement points, while the error bars are calculated as half the difference between maximum and minimum values. We estimate values of d$_{33}$ of $\sim$8.9 pmV$^{-1}$, $\sim$18.7 pmV$^{-1}$, and $\sim$15.4 pmV$^{-1}$ for C - f,  ChNCs - f, and CsNCs - f, respectively. Notably, the introduction of NCs leads to a $\sim$ two-fold enhancement in the piezoelectric coefficient value. This points to an increased crystallinity and enhanced piezoelectric properties.

We stress that the observed values for pure chitosan films are in line with those reported in previous literature, further validating our measurements. The same methodology was also used with a sample of gold (Au) that is known to be non-piezoelectric thus serving as a reference. Figure \ref{fig:FigS3} shows the displacement values at $V_{DC}$ = $V_{EL}$ for Au. The d$_{33}$ value extracted for Au is approximately 1.6 pmV$^{-1}$, which is close to the detection limit of the PFM instrument. Au can therefore be considered non piezoelectric, confirming the validity and effectiveness of our methodology. The calibration process also involved applying the previously described extraction method to a lithium niobate standard sample. This was done by adjusting the tip's sensitivity to match the datasheet values of d$_{33}$ for lithium niobate. The results of this procedure are shown in Figure \ref{fig:FigS4}. 

\section{\label{End}Conclusions}

The present study presented an innovative approach to developing piezoelectric bio-elastomers. This is achieved by enhancing the crystallinity of chitosan thin films through the incorporation of either chitin or chitosan nanocrystals into the chitosan matrix. We observed a more-than-two-fold increase in the values of the d$_{33}$ coefficient (up to 18.7 pmV$^{-1}$ compared to pure chitosan films 8.9 pmV$^{-1}$) when chitin or chitosan nanocrystals are incorporated in the films. Engineering efforts could lead to further enhancements of the crystalline structure, thereby improving the piezoelectric properties of the films. In this work, we also showed a universal method for measuring the piezoelectric properties of soft films. This is achieved by separating the piezoelectric and electrostatic response via PFM. Additionally, the piezoelectric characteristics demonstrated by the elastomer films at the mesoscopic scale could be harnessed to generate voltage from biological deformations at larger scales. When combined with optical transparency, biocompatibility, and the capacity for large-scale production through biowaste upcycling, these findings suggest the potential of chitosan-based elastomers for crafting sustainable-by-design transducers and sensors. This opens avenues for prosthetics, soft robotics, advanced human-machine interfaces, and applications in the Internet of Things. 

\section{\label{Exp}Experimental Section}

\textit{Materials}: chitosan (High MW), glycerol (MW 92.09 gmol$^{-1}$), chitin flakes, and acetic acid (99 w\%) were purchased from Sigma Aldrich. Deionized water was employed in all experiments.

\textit{Preparation of chitin nanocrystals (ChNCs) and chitosan nanocrystals (CsNCs)}: chitin and chitosan nanocrystals were synthesized as reported in the literature with minor modifications\citep{MassariUR}.  Chitin flakes (35 mg/ml) were mixed with a 3м HCl solution and constantly mixed (100 rpm) with a mechanical head stirred for 4 hours under reflux. The resulting mixture was cooled, centrifugated (4700 rpm, 15 min), filtered and rinsed in  water (3×). A colloidal suspension was formed by sonication (40\% amplitude for 10 mins by means of a Branson Digital Sonifier equipped with a 20 kHz horn tip probe). After sonication, the mixture was diluted in water and a 1м NaOH solution was added to precipitate the ChNCs. The ChNCs were then centrifuged and washed with water to neutral pH. The amount of ChNCs in the resulting gel was determined by drying the sample at 100°C under vacuum for 24 hours. A concentration equal to 11.7\% was detected. To obtain the CsNCs, ChNCs (16 mg/ml) were dispersed in a 12.5м NaOH solution and refluxed under constant stirring (100 rpm) for 12 hours. The resulting mixture was cooled, and the CsNCs were collected and purified  as above described. The amount of CsNCs in the gel was determined gravimetrically as for ChNCs. A concentration equal to 15\% was detected. 

\textit{Preparation of thin films}: three different compositions were selected: pure chitosan films, nanocomposite films comprising chitosan (40 wt\%), glycerol (40 wt\%), and chitin or chitosan nanocrystals (20 wt\%). Pure chitosan solutions were prepared by dissolving chitosan (250 mg) in 25 ml of aqueous acetic acid (AC) (1\% v/v). Solutions to produce composite films were prepared by combining 200 mg of chitosan, 200 mg of glycerol, and  the specific amount of either hydrated ChNCs (855 mg) or hydrated CsNCs(667 mg) required to reach a 20\% concentration in 25 mL of aqueous acetic acid (AC) (1\% v/v). The films were all prepared using the solvent casting method with an automatic applicator (Model: SAFA-219, S.A.M.A. Italia). The solutions were stirred, sonicated in an ultrasound bath (30 min) and then cast onto substrates placed on the chuck of the film applicator. In the case of films that underwent XRD analysis, a Kapton sheet was used as the casting substrate to facilitate the subsequent peeling of the dry film. The films analysed with AFM/PFM, were directly cast onto conductive (2.5 $\times$ 3 cm$^2$) slides made of glass coated with a thin layer of gold (100 nm). In this case, the film was cast on approximately ¾ of the entire surface of the slide, ensuring that a portion of the gold remained free and available to provide the necessary electrical contact for the PFM analysis. 

\textit{XRD analysis}: XRD patterns were recorded using an Empyrean Series 3 diffractometer (Malvern Panalytical, Malvern, United Kingdom). The configuration used is Reflection-Transmission Spinner 3.0. The intended wavelength used is K$\alpha$1 (Å): 1.540598. The X-ray tube voltage and current were 40 kV and 40 mA, respectively with a focus of 12 mm of length and 0.4 mm of width. The scan analysis was performed in continuous mode in the range [5°-40°].

\textit{Tensile tests}: stress-strain analysis was carried out on rectangular specimens prepared by an automatic cutter by means of an MTS Insight machine equipped with a 100 N load cell. Gauge length and sample width of 20 mm and 10 mm were respectively used, and the crosshead speed was set to 0.5 mm/s. The Young modulus (E) was calculated from the first linear segment of the curve. 

\textit{PFM Analysis}: PFM measurements were conducted using a Bruker Dimension Icon instrument (Bruker Corporation, Billerica, Massachusetts, U.S.A) equipped with a Platinum-Iridium coated conductive silicon probe (SCM-PIC-V2, Bruker Corporation, Billerica, Massachusetts, U.S.A). The probe had a nominal spring constant $k$ of 0.1 N/m, a nominal length of 450 µm, and a free resonance frequency of approximately 10 KHz. It was operated in contact mode. An alternating voltage of 3 V was applied to the sample during scanning to capture Amplitude and Phase images. The piezoelectric coefficient was determined by calibrating the tip with a Lithium Niobate sample. More specifically, the calibration process consisted of applying the extraction method described below to a standard sample by calibrating the sensitivity of the tip against the datasheet values of d$_{33}$ lithium niobate. The results of this operation are shown in Figure \ref{fig:FigS4}.

\textit{Details of the d$_{33}$ extraction method}: five points have been measured for each sample. Each point has been biased at various $V_{DC}$ and $V_{AC}$ following these steps:
\begin{enumerate}

\item The displacement versus frequency curve $Z-f$ is registered for each $V_{DC}$ in the range [-1.75, 1.75] V at 0.25 V step (11 $V_{DC}$ values) and the maximum of the $Z-f$ curve max\{$Z-f$\} and the quality factor has been extracted after interpolation;
\item The max\{$Z-f$\} values are plotted against the corresponding $V_{DC}$ after being normalized for an average quality factor $Q$;
\item Points (1) and (2) are repeated for the various $V_{AC}$ in the range 0.5 V – 3 V with a step of 0.25 V;
\item The electrostatic voltage $V_{EL}$ and the corresponding $Z$ are extracted from the linear fitting of the $Z$ versus $V_{DC}$ for each $V_{AC}$. Here $V_{AC}$ = 2.75 V and 3 V are neglected because of non-linear effects (see Equation \ref{eq:Eq1}); 
\item The $Z$ values extracted at point (4) are plotted against $V_{AC}$. 
\item The procedure 1-4 is repeated for each point (5 points for each kind of film) and are linearly interpolated to find the d$_{33}$ coefficient (slope for the fitting line).  
\end{enumerate}

\section{\label{Ackn}Acknowledgements}
This work has been funded by the European Union - Next Generation EU as part of the PRIN 2022 project “Biodegradable thin film electronics for massively deployable and sustainable Internet of Things applications” (2022L4YZS4), the PRIN PNRR 2022 project “Continuous THERmal monitoring with wearable mid-InfraRed sensors” (P2022AHXE5), and has received funding from the European Union - NextGenerationEU, in the framework of the iNEST – Interconnected Nord-Est Innovation Ecosystem (iNEST ECS$\_$00000043 – CUP H43C22000540006). The views and opinions expressed are solely those of the authors and do not necessarily reflect those of the European Union, nor can the European Union be held responsible for them. This work has also been funded by the 2014-2020 Research and Innovation NOP Action IV.4 – project “Biodegradable electronic packaging and sensors from biowaste upgrading” with the support of B4Plastics IQ Parklaan 2A, 3650 Dilsen-Stokkem, Belgium. Dr Domenico De Fazio also acknowledges SPIN funding from the Ca’ Foscari University of Venice.

\pagebreak
\widetext
\begin{center}
\textbf{\large Supplemental Materials: Low Dark-Current Readout for Graphene-Quantum Dots Hybrid Photodetectors}
\end{center}

\setcounter{equation}{0}
\setcounter{figure}{0}
\setcounter{table}{0}
\setcounter{page}{1}
\makeatletter
\renewcommand{\theequation}{S\arabic{equation}}
\renewcommand{\thefigure}{S\arabic{figure}}
\renewcommand{\bibnumfmt}[1]{[S#1]}
\renewcommand{\citenumfont}[1]{S#1}
\newcounter{SIfig}
\renewcommand{\theSIfig}{S\arabic{SIfig}}


\section{\label{Optical}Measurements of optical absorption of chitosan films}

Methodology for measurements of the optical absorption: the tests were carried out on chitosan films produced by a solvent casting process (starting from 1 ml of aqueous acetic acid (AC) (1\% v/v) solution with a chitosan concentration of 25 mg/ml) using an applicator film by casting the solution directly onto the sample holder. The resulting film is approximately 20 µm thick. The UV-Vis Spectrophotometry tests were carried out with an Agilent Cary 100 UV/Vis Spectrophotometer. The calibration process was carried out using the blank sample holder as a reference, \textit{i.e.} without the film deposited on it.\citep{SoleIJBM2022,KumirskaMD2010,AbdolrahimiCJP2018}

\begin{figure}[htbp!]
\centerline{\includegraphics[width=90mm]{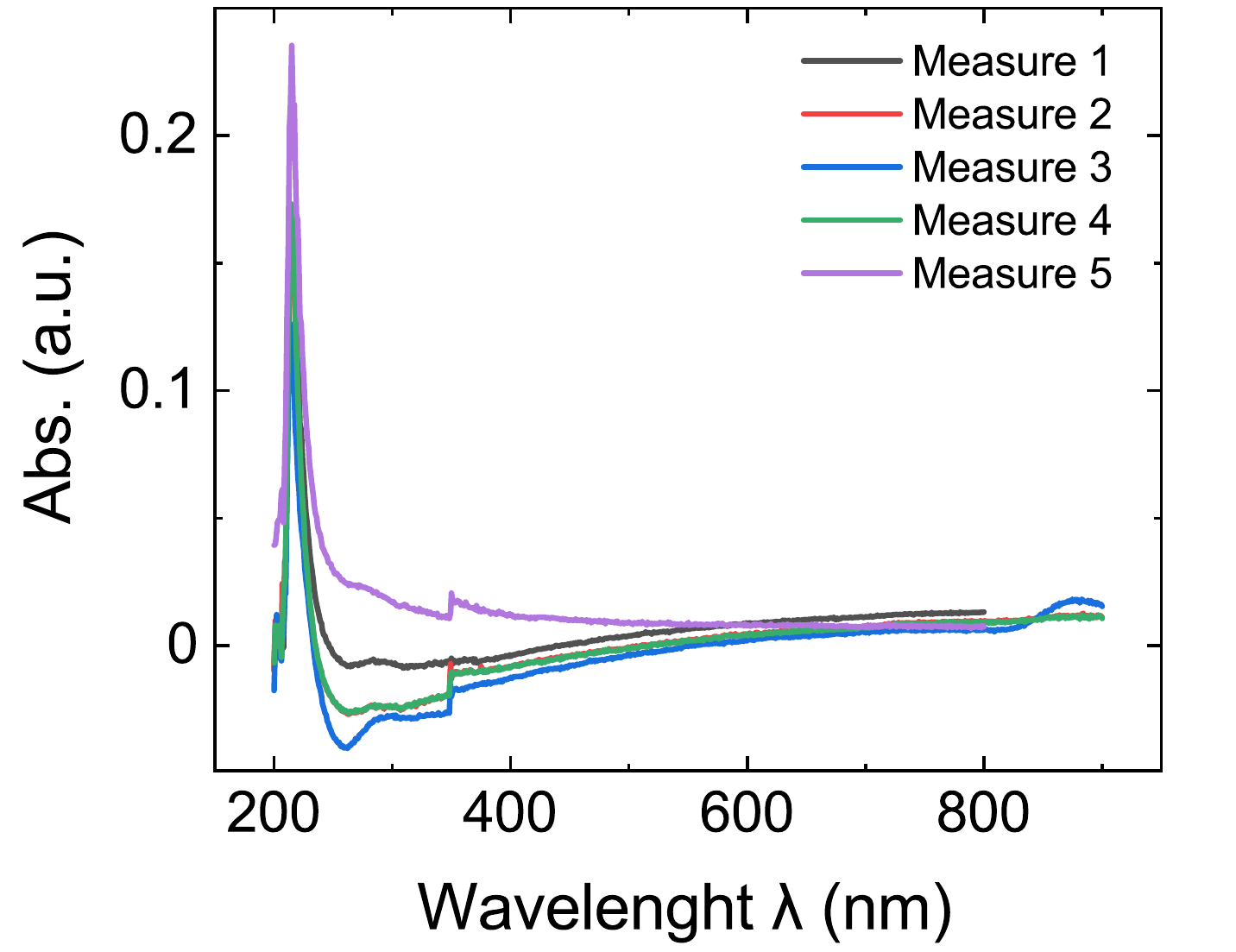}}
\caption{Optical absorption of chitosan films. The graph refers to five repeated measurements of the same sample. All curves appear flat in the visible range (wavelength between 400 - 800 nm) and with a constant value of approximately zero. It is also possible to observe an absorption peak around 200 nm, which is consistent with other results from studies on chitosan films in the literature.}
\refstepcounter{SIfig}\label{fig:FigS1}
\end{figure}

\section{\label{AFM}Atomic Force Microscopy (AFM) analysis}

\begin{figure}[H]
\centerline{\includegraphics[width=140mm]{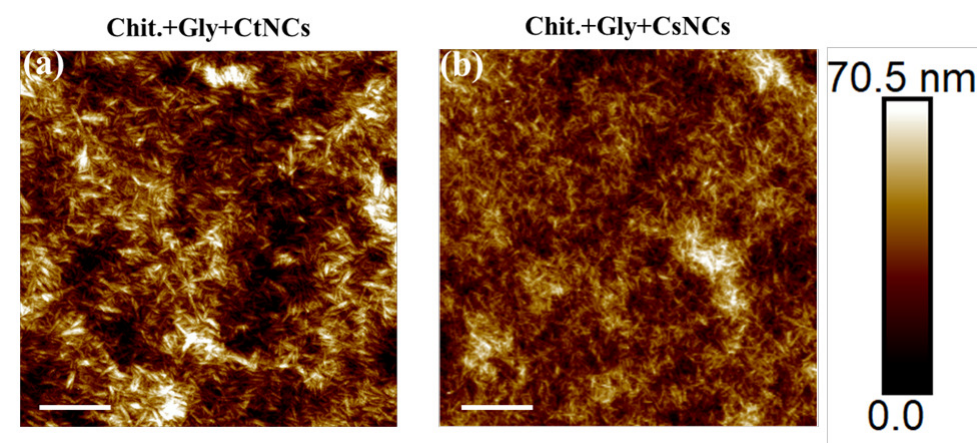}}
\caption{AFM images of topography (a-b) of chitosan + glycerol + CtNCs, and chitosan + glycerol + CsNCs films. The scan size is 10 $\times$ 10 µm$^2$ and the scale bar is 2 µm in all plots, while the colour scales on the right refer to both plots. The measurements were conducted using a Bruker Dimension Icon instrument (Bruker Corporation, Billerica, Massachusetts, U.S.A) equipped with a silicon tip on silicon nitride cantilever (SCM-PIC-V2, Bruker Corporation, Billerica, Massachusetts, U.S.A). The probe had a nominal spring constant $k$ of 0.4 N/m, a nominal length of 115 µm, and a free resonance frequency of approximately 70 KHz. It was operated in tapping mode.}
\refstepcounter{SIfig}\label{fig:FigS2}
\end{figure}

\section{\label{d33}\lowercase{d}$_{33}$ extraction}

\subsection{Measurements and d$_{33}$ extraction of gold sample}

\begin{figure}[H]
\centerline{\includegraphics[width=180mm]{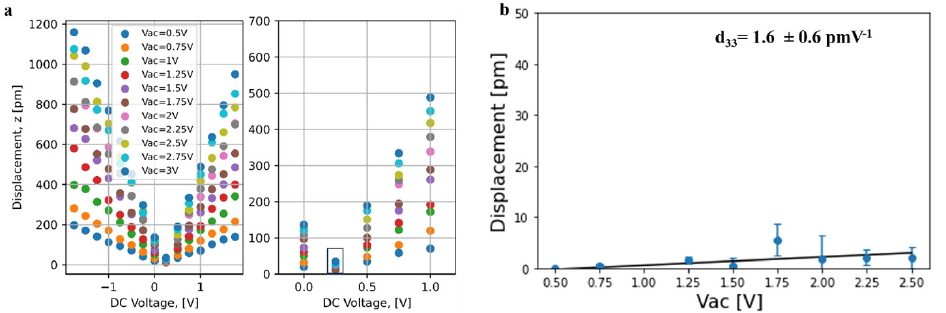}}
\caption{PFM Analysis of a gold sample: a) displacement as function of $V_{DC}$ for various $V_{AC}$. The experiments followed the methodology described in “Methods and Materials” and applied for the analysis of the chitosan-based films. The analysis shows a d$_{33}$ that is about 1.6 pmV$^{-1}$ which is close to the limit of detection of the PFM instrument. The results confirm the non-piezoelectricity of Au and the effectiveness of our methodology.}
\refstepcounter{SIfig}\label{fig:FigS3}
\end{figure}

\subsection{Measurements and d$_{33}$ extraction of lithium niobate sample}

\begin{figure}[H]
\centerline{\includegraphics[width=180mm]{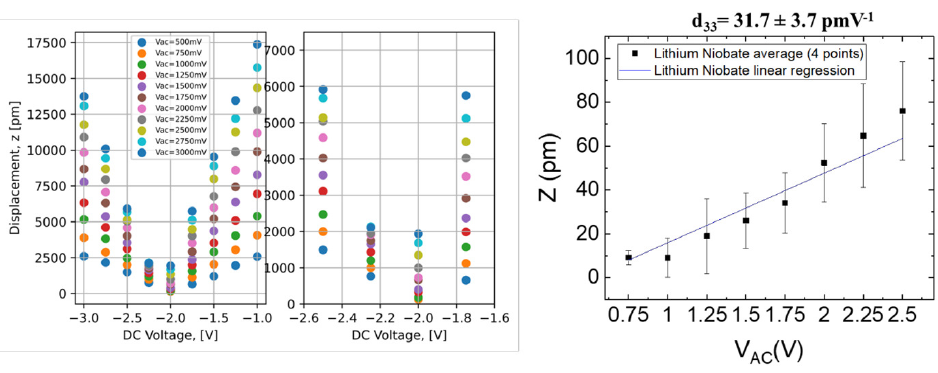}}
\caption{PFM Analysis of a lithium niobate sample: a) displacement as function of $V_{DC}$ for various $V_{AC}$. The measurements on the lithium niobate are used to calibrate the sensitivity of the tip by matching the found experimental value of d$_{33}$ with the one of the datasheet.}
\refstepcounter{SIfig}\label{fig:FigS4}
\end{figure}

\newpage

\section{\label{Tables}Benchmark tables}
\begin{table}[ht!]
\setlength{\arrayrulewidth}{0.2mm} 
\begin{tabularx}{\textwidth}{|>{\hsize=.25\hsize}X|>{\hsize=.15\hsize}X|>{\hsize=.2\hsize}X|>{\hsize=.3\hsize}X|>{\hsize=.1\hsize}X|}
\hline
\textbf{Material} & \textbf{d$_{33}$ (pmV$^{-1}$)} & \textbf{Measurement Technique} & \textbf{Processing} & \textbf{Reference}\\
\hline
\hline
Silicone rubber + Barium titanate & $22.5$ & Acquisition Voltage circuit & Mixing, moulding polarization and curing. & Ref.\citenum{NicoliniSMS2023}\\
\hline
Polyacrylonite & $40$ & - & - & Ref.\citenum{NicoliniSMS2023}\\
\hline
Thermosetting silicone + PZT & $60$ & - & - & Ref.\citenum{NicoliniSMS2023}\\
\hline
PDMS + PVDF & $8$ & - & - & Ref.\citenum{NicoliniSMS2023}\\
\hline
PDMS + poly(siloxane-imide)  & $0.4$ to 1.7 (depending on PI \%) & PFM & Polycondensation and chemical imidization & Ref.\citenum{SupplStiubianuM2022}\\
\hline
PDMS + PI & $0.5$ (pure PDMS); $17.6-26$ (depending on PI\%) & PFM & Polycondensation & Ref.\citenum{SupplTuguiJMCC2017}\\
\hline
PVDF + polyacrylonitrile & $63$ & Dynamic mechanical tester & Radical polymerization and freeze-drying processes & Ref.\citenum{SupplGuanACSAMI2023}\\
\hline
Chloroprene rubber + 60 vol\% PZT & $244$ & - & - & Ref.\citenum{OwusuAMT2023}\\
\hline
Poly(acrylonitrile butadiene) rubber + 80 vol\% PMN & $33$ & - & - & Ref.\citenum{OwusuAMT2023}\\
\hline
PDMS + 43 vol \% PZT & $6.8$ & - & - & Ref.\citenum{OwusuAMT2023}\\
\hline
PDMS + 8 vol \% BCZT & $31$ & - & - & Ref.\citenum{OwusuAMT2023}\\
\hline
PDMS + 50 vol \% PZT & $101$ & - & - & Ref.\citenum{OwusuAMT2023}\\
\hline
PDMS + 20-60 vol \% PZT\% PMN & $80-240$ & - & - & Ref.\citenum{OwusuAMT2023}\\
\hline
CNA-doped polyurethane foams & $244$ & - & - & Ref.\citenum{OwusuAMT2023}\\
\hline 
PDMS + 33–40 wt\% poly [(MMA)-co-(DR1-MA)] \% PZT\% PMN & $27$ & - & - & Ref.\citenum{OwusuAMT2023}\\
\hline 
PDMS + 30 wt\% polar polynorbornene & $37$ & - & - & Ref.\citenum{OwusuAMT2023}\\
\hline 
Polar smectic bent-core liquid crystal material confined in a biocompatible triblock copolymer (BCLC/SIBSTAR composite) & $1000$ & Mirau interferometry & 0.5 and 1 mm thick elastomer films were made by solvent casting followed by compression molding under 50 atm pressure that was released at 60 $\mu$C, when the liquid crystal component was in the X-phase. & Ref.\citenum{CharifRSCA2013}\\
\hline  
\end{tabularx}
\caption{\label{tab:Tab1} Manufacturing technology, measuring piezoelectric technique and piezoelectric coefficient d$_{33}$ values for different soft polymers}
\end{table}

\begin{table}[ht!]
\setlength{\arrayrulewidth}{0.2mm} 
\begin{tabularx}{\textwidth}{|>{\hsize=.25\hsize}X|>{\hsize=.15\hsize}X|>{\hsize=.2\hsize}X|>{\hsize=.3\hsize}X|>{\hsize=.1\hsize}X|}
\hline
\textbf{Material} & \textbf{Piezoelectric coefficient (pmV$^{-1}$)} & \textbf{Measurement Technique} & \textbf{Processing} & \textbf{Reference}\\
\hline
\hline
Diphenylalanine (Peptide) & d$_{15}$=80 & PFM & Self-assembly nanotube & Ref.\citenum{SupplKholkinACSN2010}\\
\hline
Collagen Nanofibrils (Protein) & d$_{33}$=22 & Piezotest, PM300 & Self-alignment nanofibrils & Ref.\citenum{SupplSunSCPMA2020}\\
\hline
Collagen (Protein) & d$_{14}$=12 & PFM & Use of fibrillary rat tail collagen & Ref.\citenum{SupplAliAHM2023}\\
\hline
Cellulose (Polysaccharides) & d$_{33}$=19.3 & Oscilloscope/pressure sensor & Electrospinning & Ref.\citenum{SupplSunCRCU2022}\\
\hline
Lysozyme (Protein) & d$_{33}$=6.5 & Piezometer & Drop casting & Ref.\citenum{SupplAliAHM2023}\\
\hline
Cysteine (Amino acid) & d$_{22}$=11 & - & - & Ref.\citenum{SupplAliAHM2023}\\
\hline
Glycine (Amino acid) & d$_{33}$=4.7; d$_{16}$=178 & Piezometer & Drop casting & Ref.\citenum{SupplGuerinNM2018}\\
\hline
Silk thin film & d$_{33}$=56.7 & PFM & Spin coating & Ref.\citenum{JosephIEEESJ2017,SupplJosephIEEE2018}\\
\hline
Proline Amino acid & d$_{25}$=27.5 & Impedance/gain phase analyser & Growing of crystals & Ref.\citenum{SupplAliAHM2023,GuerinN2018}\\
\hline  
\end{tabularx}
\caption{\label{tab:Tab2} Manufacturing technology, measuring piezoelectric technique and piezoelectric coefficient d$_{33}$ values for different bio-polymers.}
\end{table}

\begin{table}[ht!]
\setlength{\arrayrulewidth}{0.2mm}
\begin{tabularx}{\textwidth}{|>{\hsize=.25\hsize}X|>{\hsize=.15\hsize}X|>{\hsize=.2\hsize}X|>{\hsize=.3\hsize}X|>{\hsize=.1\hsize}X|}
\hline
\textbf{Material} & \textbf{d$_{33}$ (pmV$^{-1}$)} & \textbf{Measurement Technique} & \textbf{Processing} & \textbf{Reference}\\
\hline
\hline
Chitosan thin film (thickness = 15 µm) & 15.6 & PFM & Solvent casting and neutralization  & Ref.\citenum{SupplDeMarzoAEE2023}\\
\hline
Chitosan film (thickness = 70 µm) & 2.6 & PFM & Drop casting & Ref.\citenum{DeMarzoEP2021}\\
\hline
Chitosan pellets (thickness = 1.3 mm) & 18.4 & PFM & Manual hydraulic pressure & Ref.\citenum{SupplPraveenRSCA2017}\\
\hline
Chitosan-poly(3-hydroxybutyrate) (CS-PHB) blend thin films (PHB at 13 wt\%) (thickness = 40 µm) & 5 & PFM & Solvent casting & Ref.\citenum{SupplToalaIJBM2023}\\
\hline
Chitosan thin film (thickness = 20 µm) & 6 & Dynamic force sensor & Solvent casting & Ref.\citenum{SupplHanninenCP2018}\\
\hline
Chitosan-glycine (1:0.8) (thickness= 38 µm) & - & Dynamic pressure using a TIRA shake & Drop casting & Ref.\citenum{SupplHosseiniACSAMI2020}\\
\hline
Chitosan - PEDOT nanofibers & - & Pressure and electrochemical system & Electrospinning and H$_2$S$_{O4}$ - treatment for recrystallization & Ref.\citenum{SupplDuJCIS2020}\\
\hline
Chitosan film  & - & In-house built setup (shaker for dynamic excitation force) & Solvent casting and neutralization & Ref.\citenum{SupplHanninenPE2016}\\
\hline
\end{tabularx}
\caption{\label{tab:Tab3} Manufacturing technology, measuring d$_{33}$ technique and piezoelectric coefficient d$_{33}$ values for different chitosan-based materials.}
\end{table}

\begin{table}[ht!]
\setlength{\arrayrulewidth}{0.2mm} 
\begin{tabularx}{\textwidth}{|X|c|c|}
\hline
\textbf{Material} & \textbf{Roughness average [nm]} & \textbf{Roughness root mean square [nm]}\\
\hline
\hline
Chitosan & 0.51 & 0.67\\
\hline
Chit + Gly + CtNCs & 5.22 & 6.49\\
\hline
Chit + Gly + CsNCs & 3.51 & 4.44\\
\hline
\end{tabularx}
\caption{\label{tab:Tab4} Roughness measurements obtained from PFM scans}
\end{table}



\begin{thebibliography}{100}
\bibitem{JosephIEEE2018} Ô J. Joseph, S. G. Singh, and S. R. K. Vanjari, Piezoelectric Micromachined Ultrasonic Transducer Using Silk Piezoelectric Thin Film, IEEE Electron Device Lett. \textbf{39}, 749 (2018).

\bibitem{SunCRCU2022} B. Sun, D. Chao, and C. Wang, Piezoelectric Nanogenerator Based on Electrospun Cellulose Acetate/Nanocellulose Crystal Composite Membranes for Energy Harvesting Application, Chem. Res. Chinese U. \textbf{38}, 1005 (2022).

\bibitem{SunSCPMA2020} Y. Sun, K. Zeng, and T. Li, Piezo-/Ferroelectric Phenomena in Biomaterials: A Brief Review of Recent Progress and Perspectives, Sci. China Phys., Mech. Astron. \textbf{63}, 278701 (2020).

\bibitem{GuerinNM2018} S. Guerin et al., Control of Piezoelectricity in Amino Acids by Supramolecular Packing, Nat. Mater. \textbf{17}, 180 (2018).

\bibitem{KholkinACSN2010} A. Kholkin, N. Amdursky, I. Bdikin, E. Gazit, and G. Rosenman, Strong Piezoelectricity in Bioinspired Peptide Nanotubes, ACS Nano \textbf{4}, 610 (2010).

\bibitem{ShinEES2015} D.-M. Shin, H. J. Han, W.-G. Kim, E. Kim, C. Kim, S. W. Hong, H. K. Kim, J.-W. Oh, and Y.-H. Hwang, Bioinspired Piezoelectric Nanogenerators Based on Vertically Aligned Phage Nanopillars, Energy Environ. Sci. \textbf{8}, 3198 (2015).

\bibitem{AthenstaedtS1982} H. Athenstaedt, H. Claussen, and D. Schaper, Epidermis of Human Skin: Pyroelectric and Piezoelectric Sensor Layer, Science \textbf{216}, 1018 (1982).

\bibitem{AliAHM2023} M. Ali, M. J. Bathaei, E. Istif, S. N. H. Karimi, and L. Beker, Biodegradable Piezoelectric Polymers: Recent Advancements in Materials and Applications, Adv. Healthc. Mater. \textbf{12}, 2300318 (2023).

\bibitem{SultanaJMCB2017} A. Sultana, S. K. Ghosh, V. Sencadas, T. Zheng, M. J. Higgins, T. R. Middya, and D. Mandal, Human Skin Interactive Self-Powered Wearable Piezoelectric Bio-e-Skin by Electrospun Poly-l-Lactic Acid Nanofibers for Non-Invasive Physiological Signal Monitoring, J. Mater. Chem. B \textbf{5}, 7352 (2017).

\bibitem{DuJCIS2020} L. Du, T. Li, F. Jin, Y. Wang, R. Li, J. Zheng, T. Wang, and Z.-Q. Feng, Design of High Conductive and Piezoelectric Poly (3,4-Ethylenedioxythiophene)/Chitosan Nanofibers for Enhancing Cellular Electrical Stimulation, J. Colloid Interface Sci. \textbf{559}, 65 (2020).

\bibitem{KamelBR2022} N. A. Kamel, Bio-Piezoelectricity: Fundamentals and Applications in Tissue Engineering and Regenerative Medicine., Biophys. Rev. \textbf{14}, 717 (2022).

\bibitem{WuMD2021} Y. Wu, Y. Ma, H. Zheng, and S. Ramakrishna, Piezoelectric Materials for Flexible and Wearable Electronics: A Review, Mater. Des. \textbf{211}, 110164 (2021).

\bibitem{LemaireSMS2018} E. Lemaire, C. Ayela, and A. Atli, Eco-Friendly Materials for Large Area Piezoelectronics: Self-Oriented Rochelle Salt in Wood, Smart Mater. Struct. \textbf{27}, 025005 (2018).

\bibitem{KarakiJJAP2007} T. Karaki, K. Yan, T. Miyamoto, and M. Adachi, Lead-Free Piezoelectric Ceramics with Large Dielectric and Piezoelectric Constants Manufactured from BaTiO$_3$ Nano-Powder, Jpn. J. Appl. Phys. \textbf{46}, L97 (2007).

\bibitem{SmithJACS2012} G. L. Smith et al., PZT-Based Piezoelectric MEMS Technology, J. Am. Ceram. Soc. \textbf{95}, 1777 (2012).

\bibitem{GaoA2017} J. Gao, D. Xue, W. Liu, C. Zhou, and X. Ren, Recent Progress on BaTiO$_3$-Based Piezoelectric Ceramics for Actuator Applications, Actuators \textbf{6}, 24 (2017).

\bibitem{FanAM1999} J. Fan, W. A. Stoll, and C. S. Lynch, Nonlinear Constitutive Behavior of Soft and Hard PZT: Experiments and Modeling, Acta Mater. \textbf{47}, 4415 (1999).

\bibitem{VandenEndeJMS2007} D. A. van den Ende, P. de Almeida, and S. van der Zwaag, Piezoelectric and Mechanical Properties of Novel Composites of PZT and a Liquid Crystalline Thermosetting Resin, J. Mater. Sci. \textbf{42}, 6417 (2007).

\bibitem{PandaJMS2009} P. K. Panda, Review: Environmental Friendly Lead-Free Piezoelectric Materials, J. Mater. Sci. \textbf{44}, 5049 (2009).

\bibitem{AhamedN2020} M. Ahamed, M. J. Akhtar, M. A. M. Khan, H. A. Alhadlaq, and A. Alshamsan, Barium Titanate (BaTiO$_3$) Nanoparticles Exert Cytotoxicity through Oxidative Stress in Human Lung Carcinoma (A549) Cells, Nanomaterials 10, 2309 (2020).

\bibitem{GuanACSAMI2023} Y. Guan et al., Soft, Super-Elastic, All-Polymer Piezoelectric Elastomer for Artificial Electronic Skin, ACS Appl. Mater. Interfaces \textbf{15}, 1736 (2023).

\bibitem{StiubianuM2022} G.-T. Stiubianu, A. Bele, A. Bargan, V. O. Potolinca, M. Asandulesa, C. Tugui, V. Tiron, C. Hamciuc, M. Dascalu, and M. Cazacu, All-Polymer Piezo-Composites for Scalable Energy Harvesting and Sensing Devices, Molecules \textbf{27}, 8524 (2022).

\bibitem{TuguiJMCC2017} C. Tugui, A. Bele, V. Tiron, E. Hamciuc, C. D. Varganici, and M. Cazacu, Dielectric Elastomers with Dual Piezo-Electrostatic Response Optimized through Chemical Design for Electromechanical Transducers, J. Mater. Chem. C \textbf{5}, 824 (2017).

\bibitem{DeMarzoAEE2023} G. de Marzo et al., Sustainable, Flexible, and Biocompatible Enhanced Piezoelectric Chitosan Thin Film for Compliant Piezosensors for Human Health, Adv. Electron. Mater. \textbf{9}, 2200069 (2023).

\bibitem{VeigaMTC2020} A. G. Veiga, F. G. de A. Dias, L. do N. Batista, M. L. M. Rocco, and M. F. Costa, Reprocessed Poly(Vinylidene Fluoride): A Comparative Approach for Mechanical Recycling Purposes, Mater. Today Commun. \textbf{25}, 101269 (2020).

\bibitem{HanninenCP2018} A. Hänninen, E. Sarlin, I. Lyyra, T. Salpavaara, M. Kellomäki, and S. Tuukkanen, Nanocellulose and Chitosan Based Films as Low Cost, Green Piezoelectric Materials, Carbohydr. Polym. \textbf{202}, 418 (2018).

\bibitem{KumarG2024} R. Kumar and S. Bera, Recent Approaches in Development of Bio-Based Artificial Piezoelectric Constructs for Biomedical Applications, Giant \textbf{17}, 100214 (2024).

\bibitem{MaschmeyerCSR2020} T. Maschmeyer, R. Luque, and M. Selva, Upgrading of Marine (Fish and Crustaceans) Biowaste for High Added-Value Molecules and Bio(Nano)-Materials, Chem. Soc. Rev. \textbf{49}, 4527 (2020).

\bibitem{XuCSR2019} C. Xu, M. Nasrollahzadeh, M. Selva, Z. Issaabadi, and R. Luque, Waste-to-Wealth: Biowaste Valorization into Valuable Bio(Nano)Materials, Chem. Soc. Rev. \textbf{48}, 4791 (2019).

\bibitem{RodriguesJFB2012} S. Rodrigues, M. Dionísio, C. R. López, and A. Grenha, Biocompatibility of Chitosan Carriers with Application in Drug Delivery, J. Funct. Biomater. \textbf{3}, 615 (2012).

\bibitem{WronskaF2023} N. Wrońska, N. Katir, M. Nowak-Lange, A. El Kadib, and K. Lisowska, Biodegradable Chitosan-Based Films as an Alternative to Plastic Packaging, Foods \textbf{12}, 3519 (2023).

\bibitem{PavinattoB2010} F. J. Pavinatto, L. Caseli, and O. N. Jr. Oliveira, Chitosan in Nanostructured Thin Films, Biomacromolecules \textbf{11}, 1897 (2010).

\bibitem{HamedTFST2016} I. Hamed, F. Özogul, and J. M. Regenstein, Industrial Applications of Crustacean By-Products (Chitin, Chitosan, and Chitooligosaccharides): A Review, Trends Food Sci. Technol. \textbf{48}, 40 (2016).

\bibitem{ShamshinaACSSC2016} J. L. Shamshina, P. S. Barber, G. Gurau, C. S. Griggs, and R. D. Rogers, Pulping of Crustacean Waste Using Ionic Liquids: To Extract or Not To Extract, ACS Sustainable Chem. Eng. \textbf{4}, 6072 (2016).

\bibitem{SultankulovB2019} B. Sultankulov, D. Berillo, K. Sultankulova, T. Tokay, and A. Saparov, Progress in the Development of Chitosan-Based Biomaterials for Tissue Engineering and Regenerative Medicine, Biomolecules \textbf{9}, 470 (2019).

\bibitem{ChenAMT2020} Y. Chen et al., Piezoelectric and Photothermal Dual Functional Film for Enhanced Dermal Wound Regeneration via Upregulation of Hsp90 and HIF-1$\alpha$, Appl. Mater. Today \textbf{20}, 100756 (2020).

\bibitem{FenIEEESJ2013} Y. W. Fen and W. M. M. Yunus, Utilization of Chitosan-Based Sensor Thin Films for the Detection of Lead Ion by Surface Plasmon Resonance Optical Sensor, IEEE Sens. J. \textbf{13}, 1413 (2013).

\bibitem{LinBC2012} S. Lin, C.-C. Chang, and C.-W. Lin, A Reversible Optical Sensor Based on Chitosan Film for the Selective Detection of Copper Ions, Biomed. Eng. \textbf{24}, 453 (2012).

\bibitem{ParkPNAS2021} J. Park et al., Quadruple Ultrasound, Photoacoustic, Optical Coherence, and Fluorescence Fusion Imaging with a Transparent Ultrasound Transducer, Proc. Natl. Acad. Sci. \textbf{118}, e1920879118 (2021).

\bibitem{FukadaJPS1975} E. Fukada and S. Sasaki, Piezoelectricity of $\alpha$-Chitin, J. Polym. Sci. \textbf{13}, 1845 (1975).

\bibitem{AhmadIOPCS2020} F. B. Ahmad, M. H. Maziati Akmal, A. Amran, and M. H. Hasni, Characterization of Chitosan from Extracted Fungal Biomass for Piezoelectric Application, IOP Conf. Ser. Mater. Sci. Eng. \textbf{778}, 012034 (2020).

\bibitem{PraveenRSCA2017} E. Praveen, S. Murugan, and K. Jayakumar, Investigations on the Existence of Piezoelectric Property of a Bio-Polymer – Chitosan and Its Application in Vibration Sensors, RSC Adv. \textbf{7}, 35490 (2017).

\bibitem{HanninenPE2016} A. Hänninen, S. Rajala, T. Salpavaara, M. Kellomäki, and S. Tuukkanen, Piezoelectric Sensitivity of a Layered Film of Chitosan and Cellulose Nanocrystals, Procedia Eng. \textbf{168}, 1176 (2016).

\bibitem{ToalaIJBM2023} C. U. Toalá, E. Prokhorov, G. L. Barcenas, M. A. H. Landaverde, J. M. Y. Limón, J. J. Gervacio-Arciniega, O. A. de Fuentes, and A. M. G. Tapia, Electrostrictive and Piezoelectrical Properties of Chitosan-Poly(3-Hydroxybutyrate) Blend Films, Int. J. Biol. Macromol. \textbf{250}, 126251 (2023).

\bibitem{HosseiniACSAMI2020} E. S. Hosseini, L. Manjakkal, D. Shakthivel, and R. Dahiya, Glycine–Chitosan-Based Flexible Biodegradable Piezoelectric Pressure Sensor, ACS Appl. Mater. Interfaces \textbf{12}, 9008 (2020).

\bibitem{QinPT2023} L. Qin, Y. Zhang, Y. Fan, and L. Li, Cellulose Nanofibril Reinforced Functional Chitosan Biocomposite Films, Polym. Test. \textbf{120}, 107964 (2023).

\bibitem{YadavP2020} M. Yadav, K. Behera, Y.-H. Chang, and F.-C. Chiu, Cellulose Nanocrystal Reinforced Chitosan Based UV Barrier Composite Films for Sustainable Packaging, Polymers \textbf{12}, 202 (2020).

\bibitem{HoqueJMCA2018} N. A. Hoque, P. Thakur, P. Biswas, Md. M. Saikh, S. Roy, B. Bagchi, S. Das, and P. P. Ray, Biowaste Crab Shell-Extracted Chitin Nanofiber-Based Superior Piezoelectric Nanogenerator, J. Mater. Chem. A \textbf{6}, 13848 (2018).

\bibitem{HuNBM2023} H. Hu et al., Stretchable Ultrasonic Arrays for the Three-Dimensional Mapping of the Modulus of Deep Tissue, Nat. Biomed. Eng. \textbf{7}, 1321 (2023).

\bibitem{ShuR2021} S. Shu et al., Active-Sensing Epidermal Stretchable Bioelectronic Patch for Noninvasive, Conformal, and Wireless Tendon Monitoring, Research \textbf{2021}, 9783432
(2021).

\bibitem{SunNBE2020} T. Sun et al., Decoding of Facial Strains via Conformable Piezoelectric Interfaces, Nat. Biomed. Eng. \textbf{4}, 954 (2020).

\bibitem{DagdevirenEML2016} C. Dagdeviren, P. Joe, O. L. Tuzman, K.-I. Park, K. J. Lee, Y. Shi, Y. Huang, and J. A. Rogers, Recent Progress in Flexible and Stretchable Piezoelectric Devices for Mechanical Energy Harvesting, Sensing and Actuation, Extreme Mech. Lett. \textbf{9}, 269 (2016).

\bibitem{DuNR2020} S. Du, N. Zhou, Y. Gao, G. Xie, H. Du, H. Jiang, L. Zhang, J. Tao, and J. Zhu, Bioinspired Hybrid Patches with Self-Adhesive Hydrogel and Piezoelectric Nanogenerator for Promoting Skin Wound Healing, Nano Res. \textbf{13}, 2525 (2020).

\bibitem{MillerNA2019} N. C. Miller, H. M. Grimm, W. S. Horne, and G. R. Hutchison, Accurate Electromechanical Characterization of Soft Molecular Monolayers Using Piezo Force Microscopy, Nanoscale Adv. \textbf{1}, 4834 (2019).

\bibitem{Gervacio-ArciniegaJAP2020} J. J. Gervacio-Arciniega, E. A. Murillo-Bracamontes, M. Toledo-Solano, J. Fuentes, J. Portelles, E. Cruz-Valeriano, M. A. Palomino-Ovando, J. A. Ramirez-Sarabia, L. Hernandez-Gonzalez, and M. P. Cruz, Discrimination of a Ferroelectric from a Non-Ferroelectric Response in PFM by Phase Analyses at the Harmonics of the Applied Vac, J. Appl. Phys. \textbf{127}, 194102 (2020).

\bibitem{MassariUR} D. Massari, M. Sgarzi, M. Gigli, C. Crestini, \textit{under review} in Adv. Sustain. Syst.

\bibitem{SegatoUR} J. Segato, R. Calmanti, G. Gnoato, E. Cavarzerani, F. Rizzolio, C. Crestini, A. Perosa, M. Gigli, M. Selva,  \textit{under review} in Adv. Sustain. Syst.

\bibitem{PereiraCP2015} A. G. B. Pereira, E. C. Muniz, and Y.-L. Hsieh, $^1$H NMR and $^1$H–$^{13}$C HSQC Surface Characterization of Chitosan–Chitin Sheath-Core Nanowhiskers, Carbohydr. Polym. \textbf{123}, 46 (2015).

\bibitem{DosSantosCR2009} Z. M. dos Santos, A. L. P. F. Caroni, M. R. Pereira, D. R. da Silva, and J. L. C. Fonseca, Determination of Deacetylation Degree of Chitosan: A Comparison between Conductometric Titration and CHN Elemental Analysis, Carbohydr. Res. \textbf{344}, 2591 (2009).

\bibitem{TrungJMCC2017} T. Q. Trung and N.-E. Lee, Materials and Devices for Transparent Stretchable Electronics, J. Mater. Chem. C \textbf{5}, 2202 (2017).

\bibitem{IoelovichJC2014} M. Ioelovich, Crystallinity and Hydrophility of Chitin and Chitosan, J. Chem \textbf{3}, 7 (2014).

\bibitem{RoblesCP2016} E. Robles, A. M. Salaberria, R. Herrera, S. C. M. Fernandes, and J. Labidi, Self-Bonded Composite Films Based on Cellulose Nanofibers and Chitin Nanocrystals as Antifungal Materials, Carbohydr. Polym. \textbf{144}, 41 (2016).

\bibitem{GoodrichB2007} J. D. Goodrich and W. T. Winter, $\alpha$-Chitin Nanocrystals Prepared from Shrimp Shells and Their Specific Surface Area Measurement, Biomacromolecules \textbf{8}, 252 (2007).

\bibitem{Callister2008} W. D. Callister and D. G. Rethwisch, Fundamentals of Materials Science and Engineering: An Integrated Approach, 3rd ed (John Wiley \& Sons, Hoboken, NJ, 2008).

\bibitem{McKeeTEPBR2011} C. T. McKee, J. A. Last, P. Russell, and C. J. Murphy, Indentation versus Tensile Measurements of Young’s Modulus for Soft Biological Tissues., Tissue Eng. Part B Rev. \textbf{17}, 155 (2011).

\bibitem{AshuriBML2020} T. Ashuri, A. Armani, R. Jalilzadeh Hamidi, T. Reasnor, S. Ahmadi, and K. Iqbal, Biomedical Soft Robots: Current Status and Perspective, Biomed. Eng. Lett. \textbf{10}, 369 (2020).

\end{thebibliography}

\newpage

\begin{thebibliography}{30}
\bibitem{SoleIJBM2022} R. Sole, C. Buranello, A. Di Michele, and V. Beghetto, Boosting Physical-Mechanical Properties of Adipic Acid/Chitosan Films by DMTMM Cross-Linking, Int. J. Biol. Macromol. \textbf{209}, 2009 (2022).

\bibitem{KumirskaMD2010} J. Kumirska, M. Czerwicka, Z. Kaczyński, A. Bychowska, K. Brzozowski, J. Thöming, and P. Stepnowski, Application of Spectroscopic Methods for Structural Analysis of Chitin and Chitosan, Mar. Drugs \textbf{8}, 1567 (2010).

\bibitem{AbdolrahimiCJP2018} M. Abdolrahimi, M. Seifi, and M. H. Ramezanzadeh, Study the Effect of Acetic Acid on Structural, Optical and Mechanical Properties of PVA/Chitosan/MWCNT Films, Chin. J. Phys. \textbf{56}, 221 (2018).

\bibitem{NicoliniSMS2023} L. Nicolini, A. Sorrentino, and D. Castagnetti, A Soft Piezoelectric Elastomer with Enhanced Piezoelastic Response, Smart Mater. Struct. \textbf{32}, 105003 (2023).

\bibitem{SupplStiubianuM2022} G.-T. Stiubianu, A. Bele, A. Bargan, V. O. Potolinca, M. Asandulesa, C. Tugui, V. Tiron, C. Hamciuc, M. Dascalu, and M. Cazacu, All-Polymer Piezo-Composites for Scalable Energy Harvesting and Sensing Devices, Molecules \textbf{27}, 8524 (2022).

\bibitem{SupplTuguiJMCC2017} C. Tugui, A. Bele, V. Tiron, E. Hamciuc, C. D. Varganici, and M. Cazacu, Dielectric Elastomers with Dual Piezo-Electrostatic Response Optimized through Chemical Design for Electromechanical Transducers, J. Mater. Chem. C \textbf{5}, 824 (2017).

\bibitem{SupplGuanACSAMI2023} Y. Guan et al., Soft, Super-Elastic, All-Polymer Piezoelectric Elastomer for Artificial Electronic Skin, ACS Appl. Mater. Interfaces \textbf{15}, 1736 (2023).

\bibitem{OwusuAMT2023} F. Owusu, T. R. Venkatesan, F. A. Nüesch, R. M. Negri, and D. M. Opris, How to Make Elastomers Piezoelectric?, Adv. Mater. Technol. \textbf{8}, 2300099 (2023).

\bibitem{CharifRSCA2013} A. C. Charif, N. Diorio, K. Fodor-Csorba, J. E. Puskás, and A. Jákli, A Piezoelectric Thermoplastic Elastomer Containing a Bent-Core Liquid Crystal, RSC Adv. \textbf{3}, 17446 (2013).

\bibitem{SupplKholkinACSN2010} A. Kholkin, N. Amdursky, I. Bdikin, E. Gazit, and G. Rosenman, Strong Piezoelectricity in Bioinspired Peptide Nanotubes, ACS Nano \textbf{4}, 610 (2010).

\bibitem{SupplSunSCPMA2020} Y. Sun, K. Zeng, and T. Li, Piezo-/Ferroelectric Phenomena in Biomaterials: A Brief Review of Recent Progress and Perspectives, Sci. China Phys., Mech. Astron. \textbf{63}, 278701 (2020).

\bibitem{SupplAliAHM2023} M. Ali, M. J. Bathaei, E. Istif, S. N. H. Karimi, and L. Beker, Biodegradable Piezoelectric Polymers: Recent Advancements in Materials and Applications, Adv. Healthc. Mater. \textbf{12}, 2300318 (2023).

\bibitem{SupplSunCRCU2022} B. Sun, D. Chao, and C. Wang, Piezoelectric Nanogenerator Based on Electrospun Cellulose Acetate/Nanocellulose Crystal Composite Membranes for Energy Harvesting Application, Chem. Res. Chinese U. \textbf{38}, 1005 (2022).

\bibitem{SupplGuerinNM2018} S. Guerin et al., Control of Piezoelectricity in Amino Acids by Supramolecular Packing, Nat. Mater. \textbf{17}, 180 (2018).

\bibitem{JosephIEEESJ2017} J. Joseph, S. G. Singh, and S. R. K. Vanjari, Leveraging Innate Piezoelectricity of Ultra-Smooth Silk Thin Films for Flexible and Wearable Sensor Applications, IEEE Sens. J. \textbf{17}, 8306 (2017).

\bibitem{SupplJosephIEEE2018} Ô J. Joseph, S. G. Singh, and S. R. K. Vanjari, Piezoelectric Micromachined Ultrasonic Transducer Using Silk Piezoelectric Thin Film, IEEE Electron Device Lett. \textbf{39}, 749 (2018).

\bibitem{GuerinN2018} S. Guerin, T. A. M. Syed, and D. Thompson, Deconstructing Collagen Piezoelectricity Using Alanine-Hydroxyproline-Glycine Building Blocks, Nanoscale \textbf{10}, 9653 (2018).

\bibitem{SupplDeMarzoAEE2023} G. de Marzo et al., Sustainable, Flexible, and Biocompatible Enhanced Piezoelectric Chitosan Thin Film for Compliant Piezosensors for Human Health, Adv. Electron. Mater. \textbf{9}, 2200069 (2023).

\bibitem{DeMarzoEP2021} G. de Marzo, D. Desmaële, L. Algieri, L. Natta, F. Guido, V. Mastronardi, M. Mariello, M. T. Todaro, F. Rizzi, and M. De Vittorio, Chitosan-Based Piezoelectric Flexible and Wearable Patch for Sensing Physiological Strain, Eng. Proc. \textbf{6}, 12 (2021).

\bibitem{SupplPraveenRSCA2017} E. Praveen, S. Murugan, and K. Jayakumar, Investigations on the Existence of Piezoelectric Property of a Bio-Polymer – Chitosan and Its Application in Vibration Sensors, RSC Adv. \textbf{7}, 35490 (2017).

\bibitem{SupplToalaIJBM2023} C. U. Toalá, E. Prokhorov, G. L. Barcenas, M. A. H. Landaverde, J. M. Y. Limón, J. J. Gervacio-Arciniega, O. A. de Fuentes, and A. M. G. Tapia, Electrostrictive and Piezoelectrical Properties of Chitosan-Poly(3-Hydroxybutyrate) Blend Films, Int. J. Biol. Macromol. \textbf{250}, 126251 (2023).

\bibitem{SupplHanninenCP2018} A. Hänninen, E. Sarlin, I. Lyyra, T. Salpavaara, M. Kellomäki, and S. Tuukkanen, Nanocellulose and Chitosan Based Films as Low Cost, Green Piezoelectric Materials, Carbohydr. Polym. \textbf{202}, 418 (2018).

\bibitem{SupplHosseiniACSAMI2020} E. S. Hosseini, L. Manjakkal, D. Shakthivel, and R. Dahiya, Glycine–Chitosan-Based Flexible Biodegradable Piezoelectric Pressure Sensor, ACS Appl. Mater. Interfaces \textbf{12}, 9008 (2020).

\bibitem{SupplDuJCIS2020} L. Du, T. Li, F. Jin, Y. Wang, R. Li, J. Zheng, T. Wang, and Z.-Q. Feng, Design of High Conductive and Piezoelectric Poly (3,4-Ethylenedioxythiophene)/Chitosan Nanofibers for Enhancing Cellular Electrical Stimulation, J. Colloid Interface Sci. \textbf{559}, 65 (2020).

\bibitem{SupplHanninenPE2016} A. Hänninen, S. Rajala, T. Salpavaara, M. Kellomäki, and S. Tuukkanen, Piezoelectric Sensitivity of a Layered Film of Chitosan and Cellulose Nanocrystals, Procedia Eng. \textbf{168}, 1176 (2016).

\end{thebibliography}
\end{document}